\documentclass[a4paper,12pt,final]{article}
\usepackage{amsmath,amsfonts,amssymb,amsthm,amstext,eucal}
\usepackage{graphicx,lscape}

\def\bea{\begin{eqnarray}} 
\def\eea{\end{eqnarray}}
\def\beann{\begin{eqnarray*}} 
\def\eeann{\end{eqnarray*}}
\def\be{\begin{equation}} 
\def\ee{\end{equation}}
\def\ba{\begin{array}} 
\def\ea{\end{array}}
\def\ben{\begin{enumerate}} 
\def\een{\end{enumerate}}

\def\4{\tilde }
\def\5{\bar }  
\def\6{\partial } 
\def\7{\hat }

\def\cL{{\cal L}}
\def\cM{{\cal M}}
\font\mybb=msbm10 at 10pt
\def\bb#1{\hbox{\mybb#1}}
\def\bR {\bb{R}}
\def\bZ {\bb{Z}}
\def\bE {\bb{E}}

\DeclareMathOperator{\dvol}{dvol}
\newcommand{\fso}{\mathfrak{so}}

\newcommand{\SO}{\mathrm{SO}}
\newcommand{\SL}{\mathrm{SL}}
\newcommand{\U}{\mathrm{U}}
\newcommand{\fl}{\mathfrak{l}}
\newcommand{\ft}{\mathfrak{t}}

\begin{document}


\begin{titlepage}
\vfill
\begin{flushright}
WIS/11/02-MAR-DPP\\
\end{flushright}

\vfill

\begin{center}
\baselineskip=16pt
{\Large\bf The geometry of null rotation identifications}
\vskip 0.3cm
{\large {\sl }}
\vskip 10.mm
{\bf ~Joan Sim\'on }\\
\vskip 1cm
{\small
The Weizmann Institute of Science, Department of Particle Physics \\
Herzl Street 2, 76100 Rehovot, Israel \\
E-mail: jsimon@weizmann.ac.il}\\ 
\end{center}
\vfill
\par
\begin{abstract}
The geometry of flat spacetime modded out by a null rotation 
(boost+rotation) is analysed. When embedding this quotient spacetime in
String/M-theory, it still preserves one half of the original supersymmetries.
Its connection with the BTZ black hole, supersymmetric dilatonic waves and 
one possible resolution of its singularity in terms of nullbranes are also 
discussed.
\end{abstract}
\vfill

\vfill

\end{titlepage}


\section{Introduction}

If string theory provides us with a quantum theory of gravity, it is natural
to study the propagation of strings in time dependent backgrounds. Some
of the easiest models one can think of are constructed as lorentzian orbifolds
of Minkowski spacetime. These have recently received considerable
attention ~\cite{gary,seiberg,seibergtalk,costa,nikita}. 
See also ~\cite{vijay} for related work. The purpose of this work is to 
analyse the geometry of Minkowski spacetime modded out by a null rotation, 
what one may call null orbifold. One can look at this quotient space as some
intermediate case between modding out by a rotation and modding out by a
boost, which is still singular, but preserves one half of the 
spacetime supersymmetries. This set up was already considered in 
~\cite{arkady}, and also in the context of twisting the N=2 string 
~\cite{olaf}.

The first, and original, motivation for the present work was provided by the 
cosmological scenario presented in ~\cite{seiberg}. They found a classical
solution to Einstein equations coupled to a single scalar field subject to 
no potential with two branches. The first branch ($t<0$) describes a
contracting universe, whereas the second one ($t>0$), an expanding universe.
They are connected through a singularity at $t=0$, where the Planck scale
vanishes and the general relativity description breaks down. When embedded
in string theory, the limit $t\to 0$ corresponds to a weak coupling
regime, and the vanishing of the Planck scale tells us that higher dimensional
operators become important as $t\to 0$. Thus, a perturbative worldsheet 
description is appropiate and reliable near the singularity. A fascinating
remark was also given in ~\cite{seiberg} in order to deal with such a 
worldsheet description. It was stated there that the d-dimensional 
configuration could be understood as simply $\bR^{d-1}\times \cM^2$ where
$\cM^2$ is the two dimensional Milne universe. The fact that $\cM^2$ can be
understood as the quotient of the interior of the past and future light
cones by the action of the group $\bZ$ generated by a boost suggests that
the two dimensional worldsheet field theory might be that 
of a lorentzian orbifold of Minkowski spacetime. This possibility has been
recently studied in ~\cite{costa,nikita}. In ~\cite{costa}, a possible
resolution of the lorentzian orbifold by adding a transverse compact
spacelike circle was also considered.

Motivated by the above construction, we shall look for a spacetime geometry
that can be thought of Minkowski spacetime modded out by the action of some
vector field acting non-trivially on time, such that the quotient manifold
remains supersymmetric (modding out by a boost breaks supersymmetry completely)
and has no closed timelike curves (CTCs). It turns out such a spacetime 
geometry is unique \footnote{The possibility of a compact timelike
direction $(\xi=\partial_t)$ or adding further transverse rotations
to $\xi_{\text{null}}$ are not considered in this paper.} : it corresponds to 
Minkowski spacetime in three dimensions modded out by a null rotation. 
A null rotation in the $x^1$ direction is infinitesimally generated by the 
Killing vector

\begin{equation*}
  \xi_{\text{null}} = B_{01} \pm R_{12}= x^0\partial_1 + x^1\partial_0 
  \pm \left(x^1\partial_2 - x^2\partial_1\right) ~,
\end{equation*}

so that it consists of a boost in the $x^1$ direction plus a rotation in the
12-plane such that both the rapidity of the boost and the angle of the rotation
have the same norm. Thus, as stressed before, this is a particular modding
in the family parametrised by $\xi = \alpha B_{01} + \beta R_{12}$.
Whenever $|\beta|>|\alpha|$, there always exists a Lorentz transformation
such that $\xi=R_{12}$, whereas when $|\alpha|>|\beta|$, this freedom under
conjugations allows us to write $\xi=B_{01}$. The norm of the Killing vector 
$\xi_{\text{null}}$ vanishes at $x^-=0$, where our quotient manifold is even 
not Hausdorff, due to the existence of fixed points $(x^-=x^1=0)$. For 
$x^-\neq 0$, the spacetime looks like a strip of length proportional to 
$|x^-|$ with boundaries satisfying non-trivial identification conditions. 
This geometry and the absence of CTCs are discussed in section 2. Notice that 
the existence of a covariantly constant null vector in this spacetime does 
not allow us to view the singularity at $x^-=0$ as formed to the future of a 
non-singular surface. In that respect, it looks closer to a singular wave 
solution ~\cite{gary}. 

The fact that our modding involves both boosts and rotations raises the
question whether the above spacetime geometry is related, in some way, with
the BTZ black hole ~\cite{BTZ,henneaux}. This is discussed in section 3,
where it is shown that under a double scaling limit, the BTZ geometry reduces 
to the null rotation geometry.

The second motivation for this work was to understand the geometry behind
the nullbrane solution found in ~\cite{paper1}. Nullbranes were obtained
by Kaluza-Klein reduction along the orbits of the Killing vector
\begin{equation*}
  \xi = R\,\partial_z + \xi_{\text{null}} ~,
\end{equation*}
where $z$ stands for a compact transverse spacelike coordinate of length
$R$. Since $|\xi|>0$ everywhere and it has no fixed points, such a spacetime 
modding constitutes one possible way of resolving the singularity found in 
the null rotation geometry. It is indeed proved that such spacetime has
no closed causal curves. This is discussed in section 4, where we also
include the singular ten dimensional configuration, the so called
dilatonic wave, whose uplift to eleven dimensions can be interpreted
as the null rotation geometry. Some comments concerning the duality 
relations among nullbranes and dilatonic waves are also considered.
We conclude with some extensions of the construction
presented in this paper to curved backgrounds, trying to emphasize
the universality of this new sector in string theory.

{\bf Note added.} When this work was being completed, we learnt of
some work in progress ~\cite{nati} in which the same model is analysed.

\section{Null rotation identifications}

Given a Minkowskian spacetime in $\text{d+1}$ dimensions, any one parameter
subgroup generated by a Killing vector $\xi$ acting on this space
\[
  \text{P} \,\to\,e^\xi\,\text{P}
\]
defines a quotient space by identifying points along its orbit. In the
present case, $\xi$ can be decomposed as
\[
  \xi = \tau + \lambda~,
\]
where $\tau\in\ft=\bR^{1,d}$ is a translation and $\lambda\in\fl=\fso(1,d)$ 
is a Lorentz transformation. By conjugating with a Lorentz transformation one 
can bring $\lambda$ to a normal form. Since we are just interested in one 
parameter subgroups acting non-trivially on time, one is left with two 
possiblities

\begin{equation*}
  \begin{aligned}
    \lambda &= B_{01}(\gamma) + R_\perp(\beta) \\
    \lambda &= N_{+1}(u) + R_\perp(\beta) ~,
  \end{aligned}
\end{equation*}

where $B_{01}(\gamma)$ is an infinitesimal boost with parameter $\gamma$ along
direction $1$, $N_{+1}(u)\equiv B_{01}(u) \pm R_{12}(u)$ is a null rotation 
with parameter $u$ in the direction 1 and $R$ stands for rotations. In the
following, we shall not consider the possibility of rotating in the
transverse directions $(\beta=0)$. 

Preservation of supersymmetry excludes boosts and selects null rotations.
This can be easily proved by studying how Killing vectors $\xi$ act on
Killing spinors $\varepsilon$ through the spinorial Lie derivative
~\cite{josegeorge}
\[
  \cL_\xi\varepsilon = \nabla_\xi \varepsilon 
  + \frac{1}{4}\partial_{[a}\xi_{b]}\Gamma^{ab}\varepsilon~.
\]

There is a second possibility for acting non-trivially on time while preserving
full supersymmetry, which is by considering a time translation, 
$\xi=\tau=\partial_t$, but since we are not willing to consider compact
time directions, we shall concentrate on null rotation identifications,
which preserve one half of the supersymmetries.

\subsection{Geometry}

One way to describe the geometry of the null rotation identifications
is by fibering the original spacetime with the orbits of the associated
Killing vector and studying the effects of the identifications in this
adapted coordinate system. The orbits are given by
\begin{equation}
  \begin{aligned}
    x^-(s) &= x_0^- \\
    x^+(s) &= x_0^+ + 2x_0^1\,s + x_0^-\,s^2 \\
    x^1(s) &= x_0^1 + x_0^-\,s
  \end{aligned}
 \label{nullorbit}
\end{equation}
where $x_0^i$ are initial conditions, $s$ is the affine parameter
of the orbit and $x^\pm = x^0\pm x^2$ are lightlike coordinates.
Notice that the line located at $x^-=x^1=0$ is a line of fixed points.

There are two reasons why to study sections of spacetime with constant
$x^-$. First of all, $\xi_{\text{null}}$ does not act on $x^-$. Furthermore,
the norm of the Killing vector, $|\xi_{\text{null}}|^2=(x^-)^2$, teaches
us to consider either $x^-\neq 0$ or $x^-=0$. This is precisely what we
shall do next.

\begin{figure}
  \begin{center}
      \includegraphics[angle=-90]{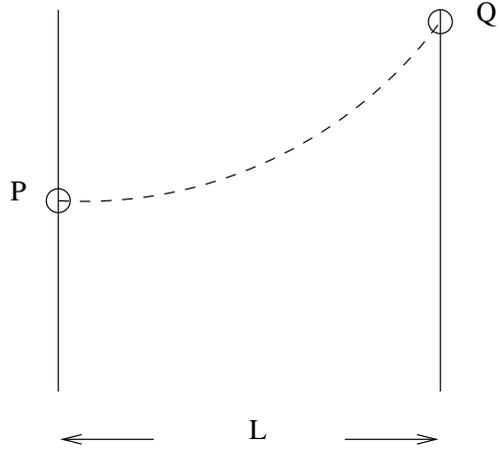} 
      \caption{\small  Geometry of the $x^- >0$ sections of flat spacetime
      after null rotation identifications. The length of the strip is
      $L=x^-\cdot\beta$. Points P and Q are identified. The dashed line
      stands for the original orbit of the Killing vector 
      $\xi_{\text{null}}$.}
    \label{geometry1}
  \end{center}
\end{figure}

\begin{figure}
  \begin{center}
      \includegraphics[angle=-90]{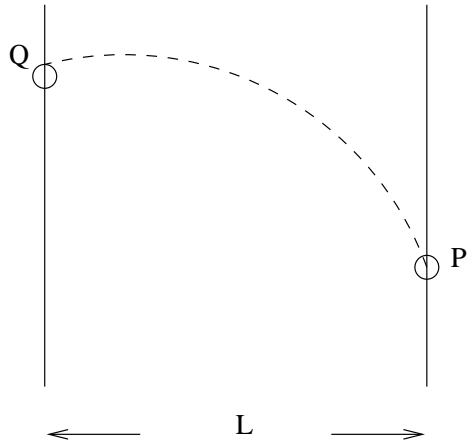}
      \caption{\small  Geometry of the $x^- < 0$ section of flat spacetime
      after null rotation identifications. The length of the strip is
      $L=|x^-|\cdot\beta$. Points P and Q are identified. The dashed line
      stands for the original orbit of the Killing vector 
      $\xi_{\text{null}}$.}
    \label{geometry2}
  \end{center}
\end{figure}

Let us start on the region where $x^-\neq 0$. The plane spanned by
$\{x^+\,,x^1\}$ is fibered by $\{s\,,x_0^+\}$, since as we vary $x_0^+$
and move along the fiber $(s)$ one covers the full subspace. We are
interested in the geometry of spacetime when non-trivial identifications
are set along the orbits \eqref{nullorbit}, that is
\[ 
  s\sim s + \beta~,
\]
for some dimensionless parameter $\beta$. The first consequence
of such an identification is that spacetime becomes a strip of length
$L=x^-\cdot\beta$ along the $x^1$ direction. We shall use the freedom to locate
the strip to set $x_0^1$ to zero. The $x^+$ direction remains a non-compact
one, but due to the parabolic nature of $x^+(s)$, the boundaries
of the forementioned strip satisfy non-trivial identification conditions.
Indeed, the point $P\equiv (x_0^+\,,0)$ is identified with
$Q\equiv (x_0^+ + L\,\beta\,,L)$
\[
  P\equiv (x_0^+\,,0) \sim (x_0^+ + L\,\beta\,,L)\equiv Q \quad 
  \forall \,x_0^+~.
\]
This geometry is illustrated in figures ~\ref{geometry1} and ~\ref{geometry2}
for $x^->0$ and $x^-<0$, respectively.

\begin{figure}
  \begin{center}
      \includegraphics[angle=-90]{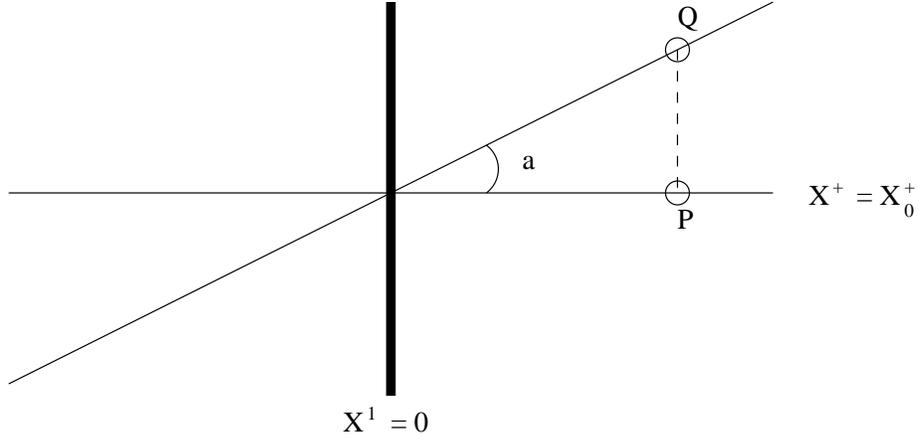}
      \caption{\small  Geometry of the $x^-=0$ section of flat spacetime
      after null rotation identifications. The bold line at $x^1=0$ stands
      for the line of fixed points. Points P and Q are identified. The angle
      $a$ is determined by $\tan\,a=2\beta$. The dashed line stands for
      a closed null curve.}
    \label{geometry3}
  \end{center}
\end{figure}

As $|x^-|$ decreases, the paraboles start degenerating
and the strip shrinks. In fact, the limit $x^-\to 0$ is singular, as can be 
seen from different points of view. First of all, the space at $x^-=0$
is not Haussdorf, due to the existence of fixed points. Its fibering is no
longer given in terms of $\{s\,,x_0^+\}$, but in terms of $\{s\,,x_0^1\}$.
This can be easily derived from the orbits \eqref{nullorbit}. Indeed, $x^1$
is left invariant at $x^-=0$, thus it remains non-compact in that
subspace. On the other hand, the previous paraboles degenerate into a line
of fixed points located at $x^1=0$. Away from $x^1=0$, one is left
with $\bR^2$ where points $P\equiv (x_0^+\,,x_0^1)$ are identified with points
$Q\equiv (x_0^+ + 2x_0^1\cdot \beta\,,x_0^1)$
\[
  P\equiv (x_0^+\,,x_0^1) \sim (x_0^+ + 2x_0^1\cdot \beta\,,x_0^1)\equiv Q 
  \quad \forall \,x_0^1\neq 0~.
\]
The geometry of the $x^-=0$ subspace is illustrated in figure 
~\ref{geometry3}.

One is thus left with two regions of spacetime ($x^->0$ and $x^-<0$) connected
through $x^-=0$, where the quotient space is not even a Haussdorf space.
In the next subsection, we shall analytically prove the existence of closed
null curves. These can easily be guessed by looking at figure ~\ref{geometry3}
and considering the null curves (dashed line on the figure),

\[
  x^-=0\quad , \quad x^1=\text{constant}
\]
connecting points P and Q which are identified by the action of the 
$\U(1)$ subgroup generated by $\xi_{\text{null}}$. Thus, at $x^-=0$, 
spacetime has also causal singularities.

At this stage, one can just remove $x^-=0$ from spacetime, leaving two 
disconnected regions with no causal singularities. A more sensible
possibility is to embed such scenario in string theory and analyse
whether the twisted sectors located at $x^-=0$ manage to resolve
such singularity. A further possibility to smooth it is to modify
the action of the $\U(1)$ subgroup by adding a transverse compact
spacelike direction
\[
  \xi=\xi_{\text{null}} \to R\partial_z + \xi_{\text{null}} ~.
\]
We postpone the discussion of such a resolution until we embed this geometry
in string theory in section 4.

\subsection{Closed causal curves}

A necessary condition for the absence of closed causal curves
in quotient spaces of the type discussed above is that the norm of the 
Killing vector generating the $\U(1)$ subgroup action is strictly positive.
Such a property is not satisfied by $\xi_{\text{null}}$, thus closed
null curves are expected at $x^-=0$, as already stressed before. The question
we would like to address is the non-existence of closed causal curves
for $x^-\neq 0$. This is non-trivial, because the forementioned criterium
is not a sufficient condition.

The proof uses an adapted coordinate system $\{t\,,x\,,y\}$ to the action
of $\xi_{\text{null}}$ ($\xi_{\text{null}}=\partial_y$) which is valid for
$x^-\neq 0$, the subspace we are interested in. This coordinate system
is defined by
\begin{equation}
  \begin{aligned}
    x^- &= t + x \\
    x^+ &= t - x + (t+x)y^2 \\
    x^1 &= (t+x)y~,
  \end{aligned}
\end{equation}
in which the three dimensional metric can be written as

\begin{equation}
  g = -(dt)^2 + (dx)^2 + (t+x)^2 (dy)^2~.
\end{equation}

We want to show there is no causal curve $x^M(\lambda)$, i.e.
$\frac{dx^M}{d\lambda}\frac{dx^N}{d\lambda}g_{MN}(x)\leq 0$ $\forall\,
\lambda$, connecting the points $(t_0\,,x_0\,,y_0)$ and
$(t_0\,,x_0\,,y_0+ \beta)$. If we assume that such a curve exists, there
must necessarily exist one value of the affine parameter $\lambda=
\lambda^\star$ where $\frac{dt}{d\lambda}$ vanishes
\[
  \exists \lambda=\lambda^\star \quad \text{s.t.} \quad
  \left(\frac{dt}{d\lambda}\right)_{\lambda^\star} = 0 ~.
\]
The norm of the tangent vector to $x^M(\lambda)$ at $\lambda=\lambda^\star$
is
\[
  \left(\frac{dx}{d\lambda}\right)_{\lambda^\star}^2 + (t+x)^2(\lambda^\star)
  \left(\frac{dy}{d\lambda}\right)_{\lambda^\star}^2 \leq 0~.
\]
Such an inequality can never be satisfied for a timelike curve, so no
closed timelike curves exist. On the other hand, the norm vanishes
when

\begin{equation*}
  \left(\frac{dx}{d\lambda}\right)_{\lambda^\star} = 
  \left(\frac{dy}{d\lambda}\right)_{\lambda^\star} = 0\,,
\end{equation*}

since we are away from $t+x=0$. Thus, at $\lambda=\lambda^\star$,
the tangent vector to the causal curve vanishes identically, which
contradicts $\lambda$ being an affine parameter. We conclude that
closed causal curves are not allowed when $t+x=x^-\neq 0$.

\section{Relation with the BTZ black hole}

The BTZ black hole line element ~\cite{BTZ} is given by
\begin{equation}
  g_{(3)} = -f^2(r)(dt)^2 + f^{-2}(r) (dr)^2 +
  r^2\left[d\phi-\frac{J}{2r^2}dt\right]^2 ~,
 \label{BTZ}
\end{equation}
the lapse function being defined by 
\begin{equation*}
  f^2(r) = -M + \left(\frac{r}{l}\right)^2 + \frac{J^2}{4r^2} ~,
\end{equation*}
where M and J are two constants of integration which are identified as the 
mass and angular momentum.

This is a solution to Einstein field equations in three dimensions with
negative cosmological constant $\Lambda$ related to the radius of curvature
through $-\Lambda = l^{-2}$. The lapse function vanishes for two values
of $r$ given by
\begin{equation}
  r_\pm = l\left[\frac{M}{2}\left\{1\pm\left[1-\left(\frac{J}{M\cdot l}
  \right)^2\right]^{1/2}\right\}\right]^{1/2} ~,
 \label{horizon}
\end{equation}
whereas $g_{tt}$ vanishes at
\begin{equation*}
  r_{\text{erg}} = l\cdot M^{1/2} ~.
\end{equation*}
These three special values of $r$ obey
\[
  r_- \leq r_+ \leq r_{\text{erg}} ~.
\]
As it happens in 3+1 dimensions for the Kerr metric, $r_+$ is the black hole
horizon, $r_{\text{erg}}$ is the surface of infinite redshift, and the region 
between $r_+$ and $r_{\text{erg}}$ is the ergosphere. The solution describes a
black hole whenever
\[
  M>0 \quad , \quad |J|\leq M\cdot l ~.
\]

Besides the continuous black hole spectrum above the vacuum (M=J=0), there is
a sort of ``bound state'' space separated from the vacuum by a mass gap of 
one unit (M=-1 and J=0), which can not be continuously deformed
to the vacuum and has neither singularities nor horizon. This is anti-de Sitter
space, which can be defined in terms of its embedding in a four-dimensional
flat space of signature $(-\,-\,+\,+)$
\[
  g_{(4)} = -(du)^2 - (dv)^2 + (dx)^2 + (dy)^2
\]
through the equation
\[
  -v^2 - u^2 + x^2 + y^2 = -l^2 ~.
\]
A system of coordinates covering the whole manifold may be introduced by
setting
\begin{equation}
  \begin{aligned}
    u &= l\cosh\mu\sin\lambda \quad , \quad v = l\cosh\mu\cos\lambda
    \quad 0\leq \mu < \infty \\
    x &= l\sinh\mu\cos\theta \quad , \quad y =  l\sinh\mu\sin\theta 
    \quad 0\leq \lambda , 2\pi
  \end{aligned}
 \label{adscoord}
\end{equation}
which yields
\begin{equation}
  g_{\text{adS}} = l^2\left[-\cosh^2\mu (d\lambda)^2 + (d\mu)^2 +
  \sinh^2\mu (d\theta)^2\right] ~.
 \label{adSmetric}
\end{equation}
Notice that strictly speaking, by an abuse of language, one refers to
anti-de Sitter space to its universal covering space, the one in which
the angular coordinate $\lambda$ has been ``unwrapped'', to avoid
the existence of closed timelike curves (CTCs).

By construction, the anti-de Sitter metric is invariant under $\SO(2,2)$.
The Killing vectors generating $\fso(2,2)$ are explicitly given below
for later reference
\begin{equation}
  \begin{aligned}
    J_{01} &= v\6_u - u\6_v \quad , \quad J_{02} = x\6_v + v\6_x ~, \\
    J_{03} &= y\6_v + v\6_y \quad , \quad J_{12} = x\6_u + u\6_x ~, \\
    J_{13} &= y\6_u + u\6_y \quad , \quad J_{23} = y\6_x - x\6_y ~. 
  \end{aligned}
 \label{adsgen}
\end{equation}

It was shown in ~\cite{henneaux} that \eqref{BTZ} can be obtained by
making identifications along the orbits generated by the Killing vector
\begin{equation}
  \xi_{\text{BTZ}} = \frac{r_+}{l}J_{12} - \frac{r_-}{l} J_{03} - J_{13} 
  + J_{23} ~.
 \label{BTZkilling}
\end{equation}
The latter contains boosts and rotations, so it is rather natural to ask 
whether there is some relation between the null rotation identification space
and the BTZ black hole. It will be proved that the former appears as a
double scaling limit of the latter \footnote{The author would like
to thank Y. Antebi and T. Volansky for discussions on this point.}.

Just by looking at the construction of both spaces, the strategy of the proof
emerges naturally. Both spaces are constructed by modding out an starting
spacetime by the action of a certain one dimensional subgroup generated by
a given Killing vector. Thus, if one establishes a precise map between the
starting configurations and the Killing vectors used in the modding, the 
relation among both quotient spaces will have been established.

The starting configurations are maximally symmetric spaces of vanishing
and negative cosmological constant, respectively. It is thus a necessary
condition, but not sufficient, to study the limit $l\to\infty$, in which
one is just left with the inside of the black hole (the exterior is pushed
away to infinity). If one further concentrates on a region pretty close
to the origin of spacetime by considering the double scaling limit
\begin{equation}
  \lambda\to \frac{t}{l} \quad , \quad \mu\to \frac{r}{l} \quad ,
  \quad l\to\infty
 \label{scaling}
\end{equation}
it is easy to check that the anti-de Sitter metric is mapped into the
Minkowski metric and the Lie algebra $\fso(2,2)$ contracts to $\fso(1,2)\ltimes
\bR^{1,2}$.

Notice that the radial coordinate $r=l\sinh\mu$ in the BTZ black hole metric
\eqref{BTZ} does not scale in the limit. This means that if $l\sim
\epsilon^{-a}$ ($a>0$), in the limit $\epsilon\to 0$
\begin{equation}
  r_\pm \quad \text{fixed} \quad , \quad M\sim\epsilon^{2a} \quad ,
  \quad J\sim\epsilon^a ~.
 \label{scaling2}
\end{equation}
Thus, the double scaling limit relating both maximally symmetric spaces
enforces the mass and angular momentum to vanish, keeping $r_\pm$ fixed.
Since the black hole interpretation disappears, $r_+$ no longer corresponds to
a horizon.

When one analyses the Killing vector \eqref{BTZkilling} in the double scaling
limit \eqref{scaling} and \eqref{scaling2}, it reduces to
\begin{equation*}
  \xi_{\text{BTZ}} \to -r_-\6_y -B_{0y} + R_{yx}~,
\end{equation*}
which is equivalent to the one used in the null rotation construction. Indeed,
just by a orientation reversal discrete transformation $(x\to -x\, , \,
y\to -y)$ and by a change of origin (conjugation under translations)
\begin{equation}
  \xi_{\text{BTZ}} \to N_{+y} = \xi_{\text{null}}~.
\end{equation}

\section{Embedding in String/M-theory}

It is straightforward to embed the previous three dimensional geometry
in string theory and M-theory. One just needs to consider the maximally
supersymmetric flat vacua of these theories and construct the quotient
space by identifying points along the orbit of $\xi_{\text{null}}$. 
As discussed in section 2, such a quotient
spacetime will be supersymmetric. In this section, we shall concentrate
on this construction for M-theory and type IIA string theory, even though
it can just as well be applied to type IIB. We shall first look for
some ten dimensional singular spacetime, whose uplift to M-theory
can be interpreted as the corresponding scenario described in section 2
when embedded in M-theory. Afterwards, we shall comment on a possible
resolution of the singularity by adding an extra compact spacelike dimension
transverse to the action of the null rotation. It will be proved that
the corresponding quotient spacetime has no closed causal curves,
and that the corresponding ten dimensional geometry is the nullbrane
supersymmetric configuration found in ~\cite{paper1}. Some comments
on duality relations among these spacetimes will be added. We conclude
with some brief discussion concerning generalizations of the previous
constructions to arbitrary curved backgrounds having an $\SO(1,2)$ isometry
subgroup.

\subsection{Dilatonic waves}

As reviewed in the introduction, one of the features of the cosmological 
scenario discussed in ~\cite{seiberg} was its higher dimensional
interpretation in terms of flat space modded out by a boost. In the same
spirit, one may look for a singular spacetime whose singularity might be
interpreted as the one discussed in section 2.  It is natural to change 
coordinates from the cartesian ones $\{x^\pm\,,x^1\}$
to the adapted ones $\{u\,,v\,,x\}$
\begin{equation}
  \begin{aligned}
    x^- &= u \\
    x^+ &= v + u\,x^2 \\
    x^1 &= u\,x
  \end{aligned}
\end{equation}
where $x$ stands for the compact coordinate along the orbit. In this
coordinate system, the metric for the three dimensional subspace of 
spacetime where $\SO(1,2)$ acts, is written as ~\cite{arkady}
\begin{equation}
  g = -du\cdot dv + u^2(dx)^2\,.
 \label{adaptmetric}
\end{equation}
The Killing vector $\xi_{\text{null}}$ becomes $\partial_x$ in the
above coordinate system and its norm vanishes at $u=0$. If one
analyses the Kaluza-Klein reduction of the eleven dimensional Minkowski
spacetime along the orbits of $\xi_{\text{null}}$, one expects a ten
dimensional configuration becoming singular precisely where the causal
singularity was located in M-theory. Proceeding in this way, one derives
a $\frac{1}{2}$-supersymmetric type IIA configuration :
\begin{equation}
  \begin{aligned}
    g & = \vert t+x \vert \left\{ds^2(\bE^8) + (dx)^2 - (dt)^2\right\} \\
    \Phi - \Phi_0 & = \frac{3}{2}\log \vert t+x \vert ~,
  \end{aligned}
\end{equation}
where we used cartesian coordinates $\{t\,,x\}$ instead of the lightlike
ones $\{u\,,v\}$.

Notice that the conformal factor $\vert t+x \vert$ does not disappear
in the Einstein frame, where the metric is given by 
\[
g_E = \vert t+x \vert^{1/4}\left\{ds^2(\bE^8) + (dx)^2 - (dt)^2\right\}\,.
\]
On the other hand, the energy-momentum tensor has the same form as the one
for an electromagnetic field $T_{mn}=ck_mk_n$. Indeed, the one form $k_{(1)}$
equals $k=d\Phi$, and since the latter defines a null one form, the
second contribution to the energy-momentum tensor, $g^{mn}\6_m\Phi\6_n\Phi$,
vanishes. Thus, $T_{mn}$ vanishes except for the components on the t-x plane
\[
T_{tt}=T_{tx}=T_{xx}=\frac{9}{4}\frac{1}{\vert t+x \vert^2}\,.
\]
Thus, as expected, the ten dimensional geometry has an essential
singularity on $t+x=0$, the lightlike hypersurface where the norm
of the Killing vector used in the reduction vanishes. Notice that even 
though the scalar curvature vanishes, both the Riemann and energy-momentum 
tensors diverge on the singularity.

\subsection{Resolution of the singularity : nullbranes}

One way of resolving the causal singularity that appears when modding out
spacetime by a null rotation is to add a compact spacelike transverse
direction in the action of the $\U(1)$ subgroup used to identify points.
In other words, one adds a transverse translation $(\tau_\perp)$ to the
null rotation Killing vector $(\xi_{\text{null}})$ and mods out spacetime
by the action of
\[
  \xi = \tau_\perp + \xi_{\text{null}}~.
\]
Such a Killing vector has an strictly positive norm, thus suggesting
the absence of closed causal curves. Furthermore, the presence of $\tau_\perp$
removes all previous fixed points left invariant by the action of 
$\xi_{\text{\null}}$. To sum up, one expects a non-singular $\frac{1}{2}$-
supersymmetric spacetime, since translations do not modify the supersymmetry
analysis presented before.

\begin{figure}
  \begin{center}
      \includegraphics[angle=-90]{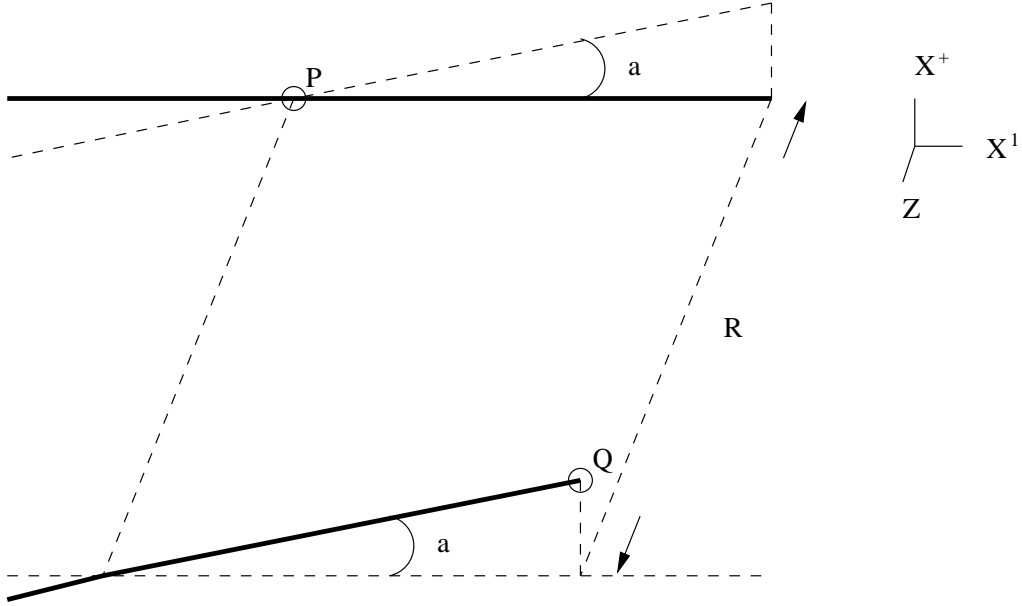}
      \caption{\small  Resolution of the singularity at $x^-=0$ by identifying
      points in spacetime through a compact translation plus a null rotation. 
      Bold lines are identified, and in particular, points P and Q are 
      identified.}
    \label{nullbranefig}
  \end{center}
\end{figure}

By analysing the orbits of $\xi$, it is straightforward to figure out the
geometry of the quotient spacetime. In figure ~\ref{nullbranefig}, it is
explicitly shown how the previous causal singularity at $x^-=0$ is resolved
by the translation $\tau_\perp$.

In order to prove that there are no closed causal curves in the quotient
spacetime, it is useful to work in adapted coordinates. These were found
in ~\cite{paper1}. Here, we just write the eleven dimensional metric
in such a coordinate system :

\begin{multline}
  g = \Lambda \left(dz + A\right)^2 + 2du\,dv - \Lambda^{-1}x^2 (du)^2 \\
  + \Lambda^{-1}(dx)^2 + 2ux \Lambda^{-1}du\,dx + ds^2(\bE^7)~,
\end{multline}
where
\begin{equation}
  \begin{aligned}
    \Lambda & = 1 + u^2 \\
    A_1 & = \Lambda^{-1}\left(u\,dx - x\,du\right)~,
  \end{aligned}
\end{equation} 
and $z$ stands for the compact transverse spacelike direction 
$(\xi=\partial_z)$.

What we want to prove is that there is no causal curve $x^M(\lambda)$,
i.e. $\frac{dx^M}{d\lambda}\frac{dx^N}{d\lambda}g_{MN}(x) \leq 0$
$\forall\,\lambda$, connecting the points $(x^i_0\,,z)$ and
$(x^i_0\,,z+R)$, because if so, that would give rise to a closed curve
in the quotient spacetime. The basic step of the proof is, once more,
to realise that for such a curve to exist, there must be some value
of the affine parameter $\lambda$ where the component of the tangent
vector $\frac{du}{d\lambda}$ vanishes. Mathematically,
\[
  \exists\, \lambda=\lambda^\star \quad \text{s.t.} \quad 
  \left(\frac{du}{d\lambda}\right)_{\lambda^\star} = 0 ~.
\]
If one evaluates the norm of the tangent vector to the {\sl assumed}
closed causal curve at $\lambda=\lambda^\star$
\[
  |\xi|^2 = \Lambda\left(\frac{dz}{d\lambda}+ \Lambda^{-1}u\frac{dx}{d\lambda}
  \right)^2_{\lambda^\star} + \Lambda^{-1}\left(\frac{dx}{d\lambda}
  \right)^2_{\lambda^\star} + \sum_{i=1}^7\left(\frac{dx^i}{d\lambda} 
  \right)^2_{\lambda^\star}~,
\]
one appreciates that it is positive or zero, so one concludes there are
certainly no closed timelike curves. Furthermore, the lower bound of the
norm is saturated if and only if
\[
  \frac{dz}{d\lambda}=\frac{dx}{d\lambda}=\frac{dx^i}{d\lambda}= 0 \quad
  \forall\,i \quad \text{at} \quad \lambda=\lambda^\star~.
\]
But such a possibility is excluded because $\lambda$ is an affine parameter.
Thus, closed null curves are also absent in this quotient spacetime.

Having resolved the singularity, we also conclude that the nullbrane
configuration discovered in ~\cite{paper1} is the ten dimensional configuration
resolving the singularity of the dilatonic wave at $u=0$.

\begin{figure}
  \begin{center}
      \includegraphics[angle=-90]{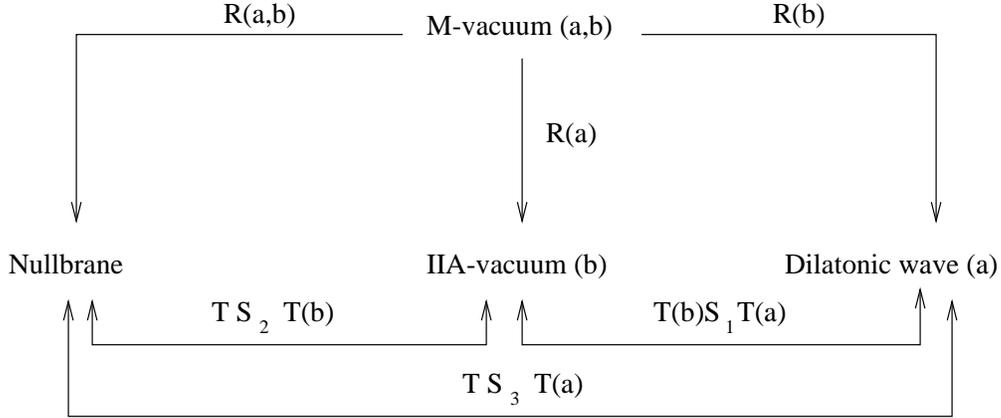}
      \caption{\small  Duality relations among nullbranes, dilatonic waves
      and flat vacua with non-trivial identifications. R stands for 
      Kaluza-Klein reduction, T for T-duality and S for an $\SL(2\,,\bZ)$ 
      transformation. $a$ and $b$ stand for the two different circles
      as explained in the text.}
    \label{dualityfig}
  \end{center}
\end{figure}

We would like to finish this subsection with some remarks concerning the
{\sl local} relation among the different configurations discussed so far.
As emphasized in ~\cite{paper1, paper2,costa1,gutperle}, any M-theory 
background with various identifications defines a moduli space of Kaluza-Klein 
reductions. Each point $P$ in this moduli corresponds to a type IIA 
configuration obtained by reduction along the orbits of a Killing vector 
field $\xi_P$. The freedom in choosing $\xi_P$ is equivalent to the freedom
in identifying the different circles in this background as the M-theory
circle. Different type 	IIA configurations obtained in this way will all
be U-dual to each other \footnote{Actually, only the full backgrounds
keeping all momentum modes are dual.}. In the case at hand, our starting 
M-theory configuration is Minkowski spacetime with two different 
identifications [M-vacuum (a,b) in figure ~\ref{dualityfig}] : along the $S^1$
circle, generated by $R\partial_z$, and along the null rotation, generated
by $\xi_{\text{null}}$. There are three different circles 
(Killing vector fields) that one can identify with the M-theory circle :

\begin{equation*}
  \begin{aligned}
    \xi_{(1)} &= R\partial_z \\
    \xi_{(2)} &= \xi_{\text{null}} \\
    \xi_{(3)} &= R\partial_z + \xi_{\text{null}}
  \end{aligned}
\end{equation*}

Reduction along the orbits generated by $\xi_{(1)}$ [R(a) in figure
~\ref{dualityfig}] gives rise to Minkowski spacetime modded out by a null 
rotation [IIA-vacuum (b) in figure ~\ref{dualityfig}]. On the other hand,
$\xi_{(2)}$ [R(b)] gives rise to a dilatonic wave with a transverse circle
[Dilatonic wave (a)] and $\xi_{(3)}$ [R(a,b)] to the nullbrane configuration. 
All of them are connected to each other through the chain of dualities 
$\text{T\,S\,T'}$, where $\text{T}$ stands for T-duality and $\text{S}$ for 
an $\SL(2\,,\bZ)$ transformation in type IIB, reminiscent of the 
$\SL(2\,,\bZ)$ symmetry of the torus spanned by 
$\{R\partial_z\,,\xi_{\text{null}}\}$. This statement can be easily
checked by starting with the nullbrane configuration and first changing
coordinates from $\{u\,,v\,,x\}$ to $\{\4u\,,\4v\,,\4x\}$

\begin{equation}
  \begin{aligned}
    \4x &= \frac{x}{u} -1 \\
    \4u &= u \\
    \4v &= v - \frac{x^2}{u}
  \end{aligned}
\end{equation}

in which the configuration is described by
\begin{equation}
  \begin{aligned}
    g &= \Lambda^{1/2}\left\{-d\4u\cdot d\4v + ds^2(\bE^7)\right\} +
    \Lambda^{-1/2}\4u^2(d\4x)^2 \\
    \Phi &= \Phi_0 + \frac{3}{4}\log \Lambda \\
    A_{(1)} &= \Lambda^{-1}\4u^2 d\4x ~,
  \end{aligned}
\end{equation}
where the scalar function $\Lambda$ is defined by $\Lambda = 1 + \4u^2$.
All the involved T-duality transformations are along the $\4x$ (or its
dual directions). On the other hand, taking into account the most
general $\SL(2\,,\bZ)$ transformation acting on $\lambda=C_{(0)} + \text{i}
e^{-\Phi}$, where $C_{(0)}$ is the RR zero form

\[
  \lambda^\prime = \frac{a\lambda + b}{c\lambda + d} \quad \text{ad-bc}=1~,
\]
one can check that under the forementioned chain of dualities, the nullbrane
configuration is mapped to :

\begin{itemize}
  \item[(i)] Minkowski spacetime with null rotation identifications
  for the choice $a=d=-c$, $b=0$ and $d^2=1$.
  \item[(ii)] Dilatonic wave in a transverse circle for the choice
  $a=c=-b$, $d=0$ and $c^2=1$.
\end{itemize}

These duality maps are illustrated in figure ~\ref{dualityfig}.

\subsection{Extension to curved backgrounds}

The purpose of the present subsection is to emphasize that the
above construction is not relying on the flatness of the starting
spacetime configuration, but on the existence of an $\SO(1,2)$
isometry. Indeed, any spacetime compatible with string theory
having an $\SO(1,2)$ group as a subgroup of its isometry group
allows such a construction. Thus, as stressed in ~\cite{paper2},
the existence of an $\SO(1,2)$ symmetry and the possibility
of identifying points along the null rotation subgroup, defines
a new sector in string theory. 

If one starts from the eleven dimensional M2-brane configuration, 
and reduces along the orbit of the Killing vector $\xi_{\text{null}}$,
one ends up with a dilatonic wave propagating on a fundamental string,
which preserves one fourth of the spacetime supersymmetries, and
is described by
 
\begin{equation}
  \begin{aligned}
    g & = \vert t+ x\vert \left\{U^{-1}[-(dt)^2 + (dx)^2] 
    + ds^2(\bE^8)\right\} \\
    \Phi - \Phi_0 & = \frac{3}{2}\log \vert t+ x\vert - \frac{1}{2}\log U \\
    H_{(3)} & = (t+x) dt\wedge dx\wedge dU^{-1}~,
  \end{aligned}
\end{equation}

where $U$ is an harmonic function defined on $\bE^8$.

Proceeding analogously with the eleven dimensional M5-brane configuration, 
one ends up with a dilatonic wave propagating on a D4-brane 

\begin{equation}
  \begin{aligned}
    g & = \vert t+ x\vert \left\{U^{-1/2}[-(dt)^2 + (dx)^2 + ds^2(\bE^2)] 
    + ds^2(\bE^5)\right\} \\
    \Phi - \Phi_0 & = \frac{3}{2}\log \vert t+ x\vert - \frac{1}{4}\log U \\
    \star G_{(4)} & = (t+x) dt\wedge dx\wedge \dvol\bE^2\wedge dU ~,
  \end{aligned}
\end{equation}

where $U$ is now defined on $\bE^5$.

If one considers the Mkk-monopole, one ends
up with a dilatonic wave propagating on a KKA-monopole 

\begin{equation}
  \begin{aligned}
    g & = \vert t+ x\vert \left\{U^{-1/2}[-(dt)^2 + (dx)^2 + ds^2(\bE^4)] 
    + ds^2_{TN}\right\} \\
    \Phi - \Phi_0 & = \frac{3}{2}\log \vert t+ x\vert 
  \end{aligned}
\end{equation}

It is straightforward to extend these constructions to any background
having an $\SO(1,2)$ isometry.

\medskip
\section*{Acknowledgments}
\noindent
The author would like to thank O. Aharony, L. \'Alvarez-Gaum\'e,
J. Barb\'on, M. Berkooz and J. M. Senovilla for discussions on the topics 
presented in this work and J. M. Figueroa-O'Farrill for many discussions
on flux and null branes. It is also a pleasure to thank the theory
division at CERN for hospitality during the intermediate stages
of the present work. This research has been supported by a Marie Curie 
Fellowship of the European Community programme ``Improving the Human 
Research Potential and the Socio-Economic knowledge Base'' under the 
contract number HPMF-CT-2000-00480.


\end{document}